\documentclass[journal]{IEEEtran}
\ifCLASSINFOpdf
\else
\fi
\pdfoutput=1
\usepackage{array}
\newcolumntype{L}{>{\centering\arraybackslash}m{1cm}}
\newcolumntype{M}{>{\centering\arraybackslash}m{3cm}}
\usepackage{mathastext}
\usepackage{chapterbib}    
\usepackage{color}         
\usepackage{graphics}      
\usepackage{array}
\usepackage{multirow}
\usepackage[pdftex]{graphicx}      
\usepackage{longtable}     
\usepackage{epsf}          
\usepackage{bm}            
\usepackage{verbatim}			
\usepackage{textcomp}
\usepackage[colorlinks=true]{hyperref}  

\begin{document}

\title{High Dynamic Range X-ray Detector Pixel Architectures Utilizing Charge Removal}

\author{Joel~T.~Weiss,
				Katherine~S.~Shanks,
				Hugh~T.~Philipp,
				Julian~Becker,
				Darol~Chamberlain,
				Prafull~Purohit,
				Mark~W.~Tate,
				Sol~M.~Gruner%

\thanks{This work has been submitted to the IEEE for possible publication. Copyright may be transferred without notice after which this version may no longer be accessible.}
\thanks{J. T. Weiss, J. Becker, D. Chamerlain, and S. M. Gruner are with the Cornell High Energy Synchrotron Source (CHESS), Cornell University, Ithaca, NY 14853-8001, USA, and the Cornell Laboratory of Atomic and Solid State Physics, Cornell University, Ithaca, NY 14853-2501, USA (email: smg26@cornell.edu).}
\thanks{K. S. Shanks, H. T. Philipp, P. Purohit, and M. W. Tate are with the Cornell Laboratory of Atomic and Solid State Physics, Cornell University, Ithaca, NY 14853-2501, USA.}}

\maketitle

\begin{abstract}
Several charge integrating CMOS pixel front-ends utilizing charge removal techniques have been fabricated to extend dynamic range for x-ray diffraction applications at synchrotron sources and x-ray free electron lasers (XFELs). The pixels described herein build on the Mixed Mode Pixel Array Detector (MM-PAD) framework, developed previously by our group to perform high dynamic range imaging. These new pixels boast several orders of magnitude improvement in maximum flux over the MM-PAD, which is capable of measuring a sustained flux in excess of 10$\bf{^{8}}$ x-rays/pixel/second while maintaining sensitivity to smaller signals, down to single x-rays. To extend dynamic range, charge is removed from the integration node of the front-end amplifier without interrupting integration. The number of times this process occurs is recorded by a digital counter in the pixel. The parameter limiting full well is thereby shifted from the size of an integration capacitor to the depth of a digital counter. The result is similar to that achieved by counting pixel array detectors, but the integrators presented here are designed to tolerate a sustained flux $>$10$\bf{^{11}}$ x-rays/pixel/second. Pixel front-end linearity was evaluated by direct current injection and results are presented. A small-scale readout ASIC utilizing these pixel architectures has been fabricated and the use of these architectures to increase single x-ray pulse dynamic range at XFELs is discussed briefly.

\end{abstract}

\IEEEpeerreviewmaketitle

\section{Introduction}

\IEEEPARstart{A}{dvances} in synchrotron radiation light source technology have opened new lines of inquiry in material science, biology, and everything in between. However, x-ray detector capabilities must advance in concert with light source technology to fully realize experimental possibilities. X-ray free electron lasers (XFELs) place particularly large demands on the capabilities of detectors, and developments towards diffraction-limited storage ring sources also necessitate detectors capable of measuring very high flux \cite{Carini2012,Nugent2010,denes2014pixel}.

For example, in coherent diffractive imaging experiments, measurement of both the intense direct beam and very low fluence wide angle scatter at the same time greatly facilitates sample reconstruction \cite{Giewekemeyer2014}. This necessitates a detector capable of measuring not only high sustained flux, but also very small signals simultaneously.

Detectors with wide dynamic ranges are needed to bridge the gap between x-ray light source technology and detector technology. Previously, our group collaboratively designed a Mixed Mode Pixel Array Detector (MM-PAD) to operate along these lines \cite{Tate2013,Schuette2008}. MM-PAD functionality will be discussed briefly. The present work describes an ASIC containing several pixel front-end test structures which build on the MM-PAD detector framework. Measurements from these test structures are presented.

\section{Dynamic Range Extension By Charge Removal}
The full well of a simple integrating pixel, defined as the amount of integrated photocurrent that can be stored in such a pixel, is limited by the size of the front-end amplifier feedback capacitance for a given output voltage swing. Increasing the integration capacitance to increase full well is ultimately constrained by pixel size. Perhaps more importantly however, a larger integration capacitance couples the output noise of the integrating amplifier to its front-end more strongly, leading to a larger equivalent noise charge, and thereby obscuring small signals. Thus increasing the well depth of an integrating pixel while maintaining sensitivity to single photon signals requires architectures more sophisticated than a simple, traditional integrator.

\begin{figure}[!t]
\centering
\includegraphics[width=2.748in]{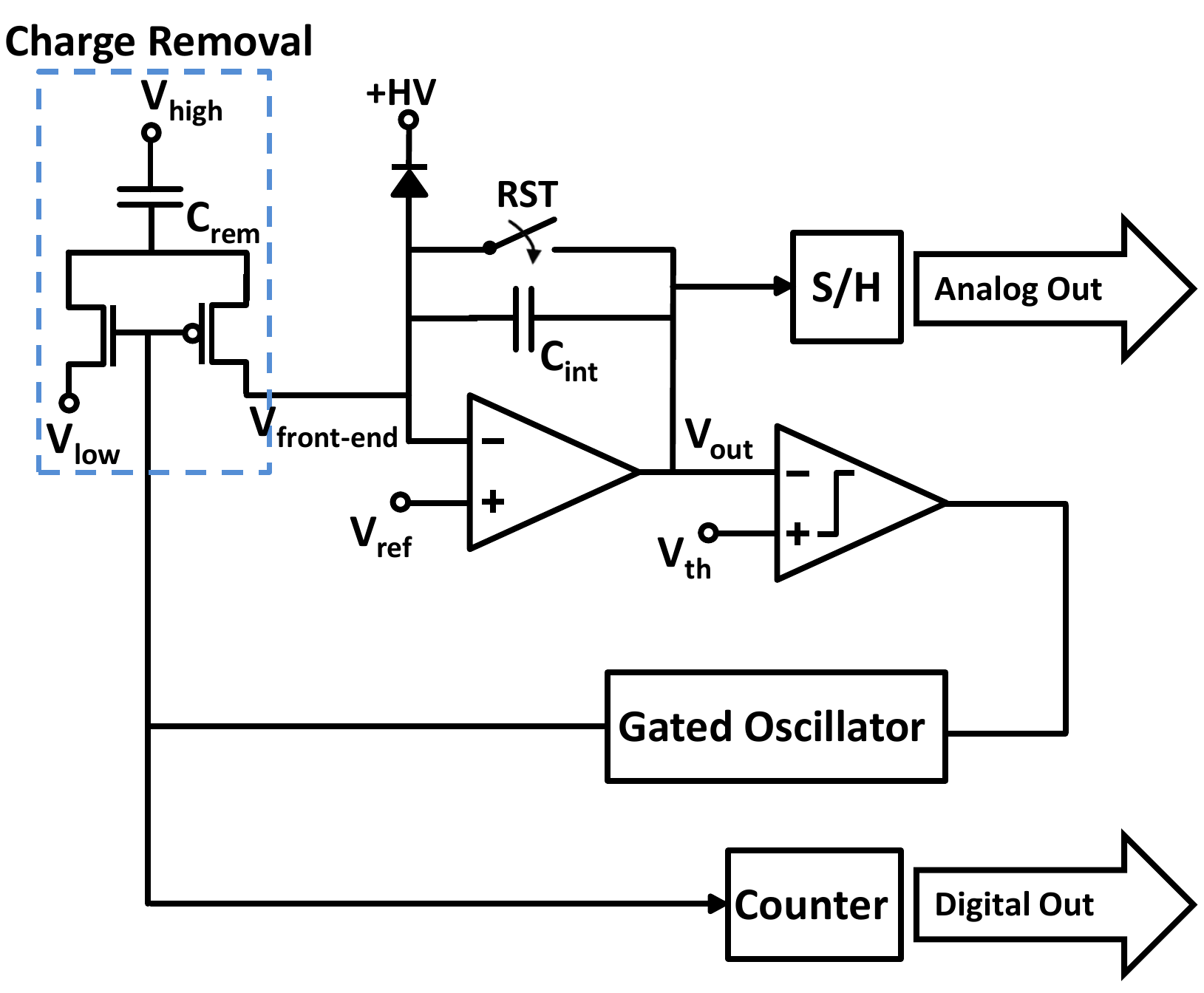}
\caption{Simplified MM-PAD schematic. The switched capacitor for charge removal is enclosed in the dotted box. In figures 1 to 4 the input results from x-rays stopped in the reverse-biased diode attached to +HV.}\label{mmpad}
\end{figure}

In the MM-PAD, photocurrent resulting from absorption of x-rays in a reverse-biased photodiode is integrated onto a charge sensitive amplifier whose output is monitored by a comparator. When the amplifier output ($V_{out}$ in figure \ref{mmpad}) crosses an externally set threshold ($V_{th}$ in figure \ref{mmpad}), a gated oscillator is enabled that triggers a counter and the switched capacitor circuit enclosed in the dotted box in figure \ref{mmpad}. With each pulse of the gated oscillator this switched capacitor removes a fixed quantity of charge ($\Delta Q = C_{rem}(V_{front-end}-V_{low})$) from the integration node while an in-pixel counter is incremented. The charge removal incurs no dead time and helps the integrator avoid saturation. The integration capacitance is sized such that the signal from a single 8 keV x-ray is readily measurable with excellent signal to noise. This strategy shifts the full well limiting parameter from the size of a capacitor to the depth of the in-pixel digital counter. The MM-PAD achieves a full well of $4x10^7$ 8 keV x-rays/pixel/frame in addition to framing at 1kHz \cite{Tate2013}.

Mixing digital and analog modes in the MM-PAD has limitations. Most prominently, in the existing design, digitization and removal of integrated charge takes approximately 500ns. The charge removed per gated oscillator cycle, $\Delta Q$, is roughly equal to the integrated photocurrent generated by 200 8 keV x-rays converted in the silicon x-ray sensor, resulting in a sustained flux capability of $~4 \textrm{x}10^8$ 8 keV x-rays/pixel/s. 

The goal of the present work is to create pixels which tolerate an even greater sustained flux. Within the MM-PAD framework, this requires increasing the maximum rate of charge removal events and increasing $\Delta Q$, the charge removed in each cycle. Three new pixel designs that accomplish this are summarized below.

\section{New Charge Removal Strategies}
Three high dynamic range pixel architectures relying on charge removal techniques were developed \cite{Shanks}. The front-ends of these pixels were laid out and fabricated in TSMC 180nm mixed-signal CMOS.

\subsection{MM-PAD 2.0}
\begin{figure}[!t]
\centering
\includegraphics[width=3.164in]{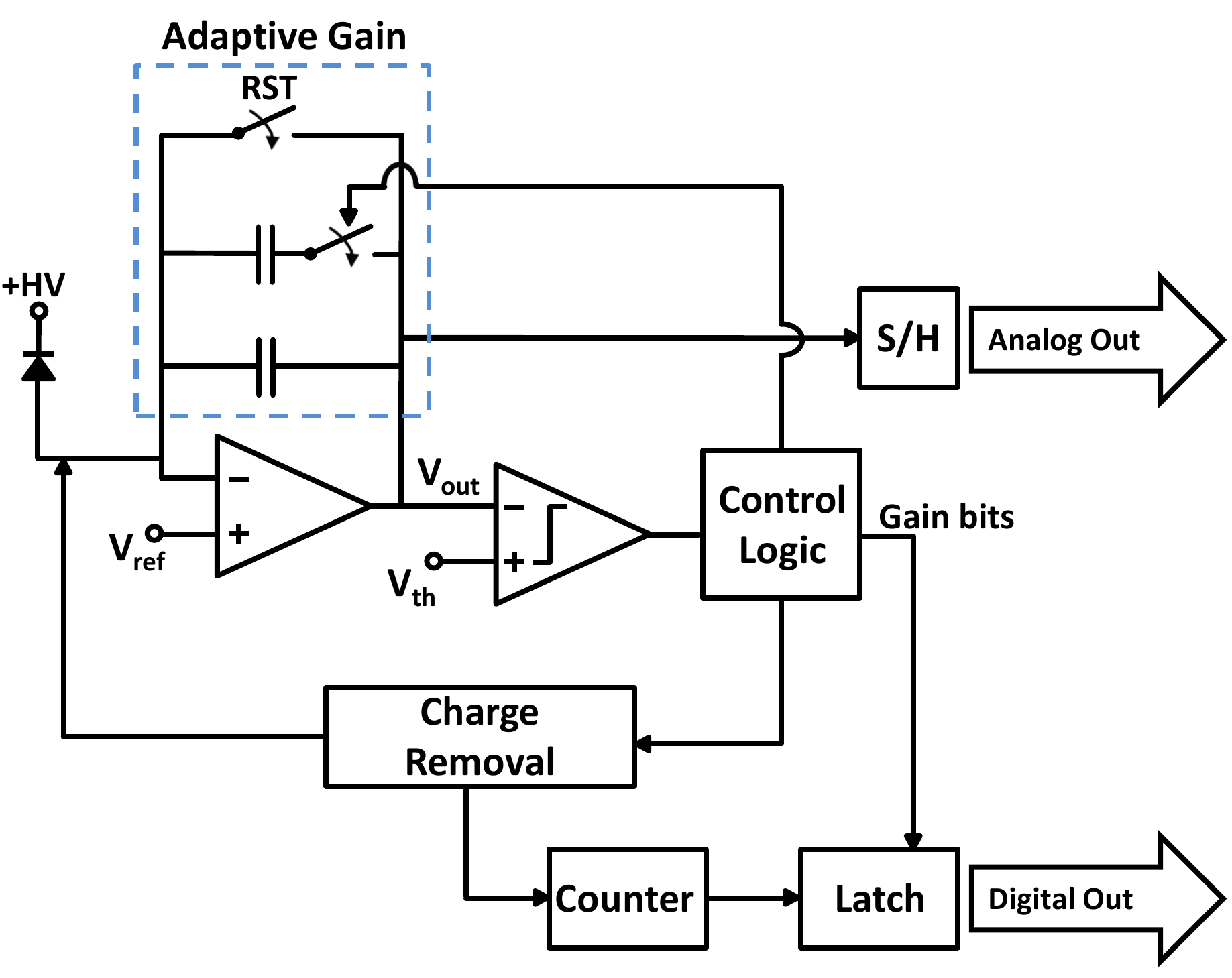}
\caption{Simplified MM-PAD 2.0 schematic. Control logic box engages adaptive gain (enclosed in the dotted box) prior to enabling switched capacitor charge removal.}\label{mmpad2}
\end{figure}

The first pixel architecture is a scaled version of the original MM-PAD and is depicted in figure \ref{mmpad2}. In contrast to the MM-PAD, the MM-PAD 2.0 incorporates adaptive gain, as demonstrated previously by detectors such as the AGIPD \cite{Henrich2011}. The MM-PAD 2.0's charge removal circuit will not trigger unless the lowest-gain stage has already been engaged. This allows the use of a larger charge removal capacitor than in the original MM-PAD, thereby increasing $\Delta Q$. In the readout ASIC discussed here, 6 combinations of total feedback capacitance and charge removal capacitance were tested. Charge removal capacitors of 1800 fF and 2630 fF (with matched feedback capacitors) exhibited incomplete charge removal at the maximum oscillator frequency. A version of the pixel with a total maximum feedback capacitance of 2630fF and a charge removal capacitance of 880fF exhibited the most robust performance. The results presented in section V are from this variant. The high-gain stage has a feedback capacitance of 40fF, small enough to resolve the signal from one 8 keV x-ray. To increase measurable sustained flux further, the maximum frequency of charge removal has been increased by a factor of 50. 

As in the original MM-PAD, the charge removal circuitry consists of a gated oscillator which toggles a switched capacitor (not explicitly shown in figure \ref{mmpad2}) to remove charge from the integration node. The MM-PAD 2.0 can tolerate larger photocurrent spikes (integrating $>10^3$ x-rays before relying on charge removal) and higher sustained photocurrent than the original MM-PAD.

\subsection{Charge Dump Oscillator (CDO)}
\begin{figure}[!t]
\centering
\includegraphics[width=3.3in]{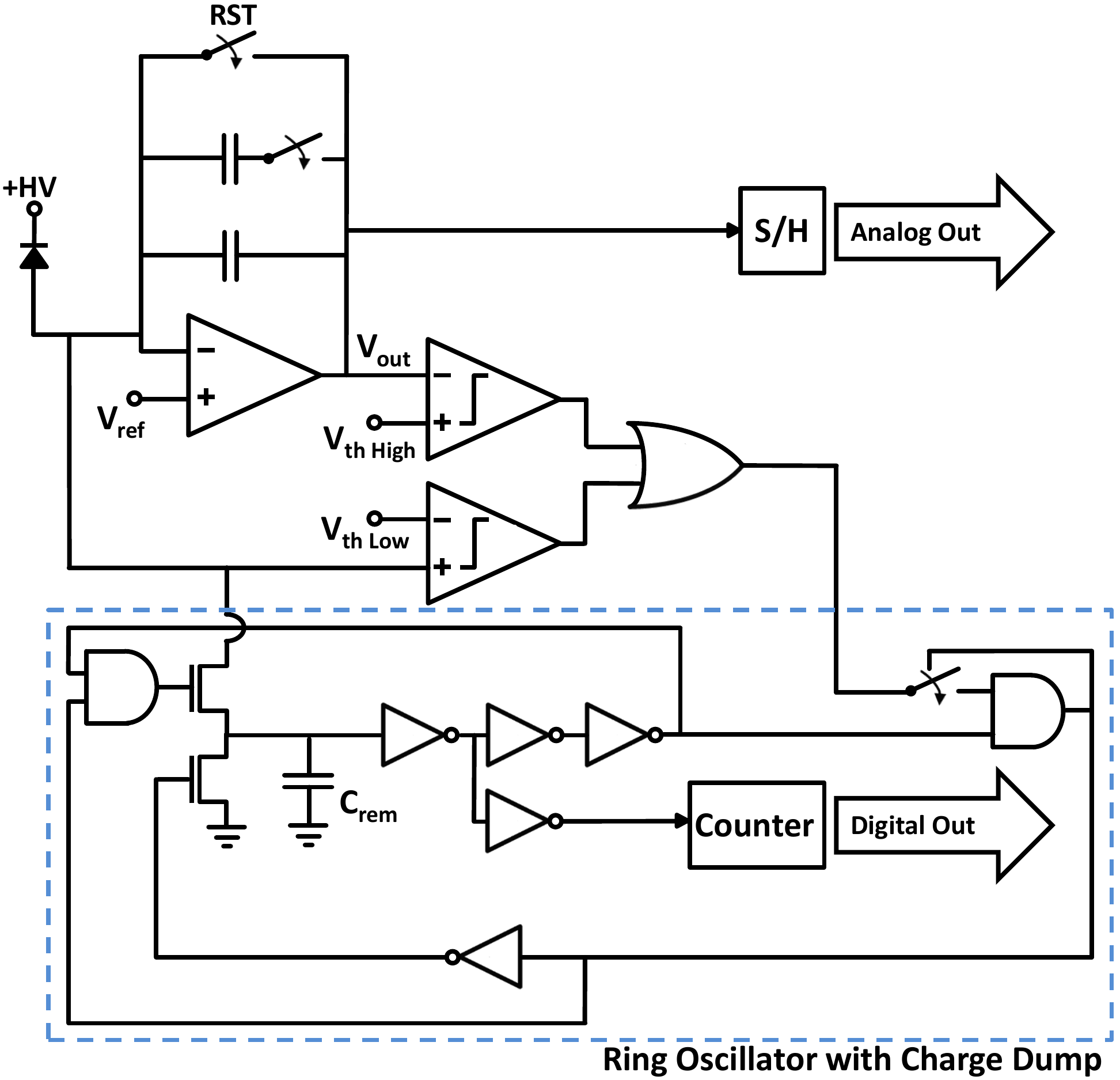}
\caption{Simplified CDO schematic. The ring oscillator charge removal circuitry is enclosed in the dotted box. Selectable gain was replaced with an adaptive gain scheme in the most recent fabrication.}\label{cdo}
\end{figure}

The CDO pixel design aims to scale the rate of charge removal with the rate of charge arrival by combining the charge removal switched capacitor with the oscillator driving it. Depicted in figure \ref{cdo}, the frequency of charge removal is set by the propagation of digital signals in the ring oscillator and the charging rate of the removal capacitor, $C_{rem}$. When a comparator indicates that a threshold has been crossed, the oscillator is activated and $C_{rem}$ connects to the pixel's integration node. Once $C_{rem}$ has charged to the switching threshold of the adjacent inverter, the capacitor is detached and the charge accumulated onto it is dumped to ground. The faster $C_{rem}$ charges while attached to the front-end, the faster charge is removed.

In addition to one comparator monitoring the integrator output, as in the MM-PAD, a second comparator monitors the pixel front-end voltage. Charge removal is also triggered by any significant deviations of this voltage from $V_{ref}$. Deviations from $V_{ref}$ indicate that the integrator is unable to keep up with incident photocurrent, in which case charge removal is needed.

At low x-ray flux, the circuit operates similarly to the MM-PAD, but the tracking of charge removal rate with incident photocurrent should allow the CDO to continuously integrate larger sustained inputs. While the inverter thresholding of the CDO is very fast, it is more susceptible to fabrication process variation than the other designs presented here. In this case, variation could result in different quantities of charge removed per cycle in each pixel. Although this can be calibrated, it is an additional complication. (Note that the schematic in figure \ref{cdo} does not show the adaptive gain circuitry that is included in the 16x16 pixel array test ASIC described in section VII).

\subsection{Capacitor Flipping Charge Removal}
\begin{figure}[!t]
\centering
\includegraphics[width=3.218in]{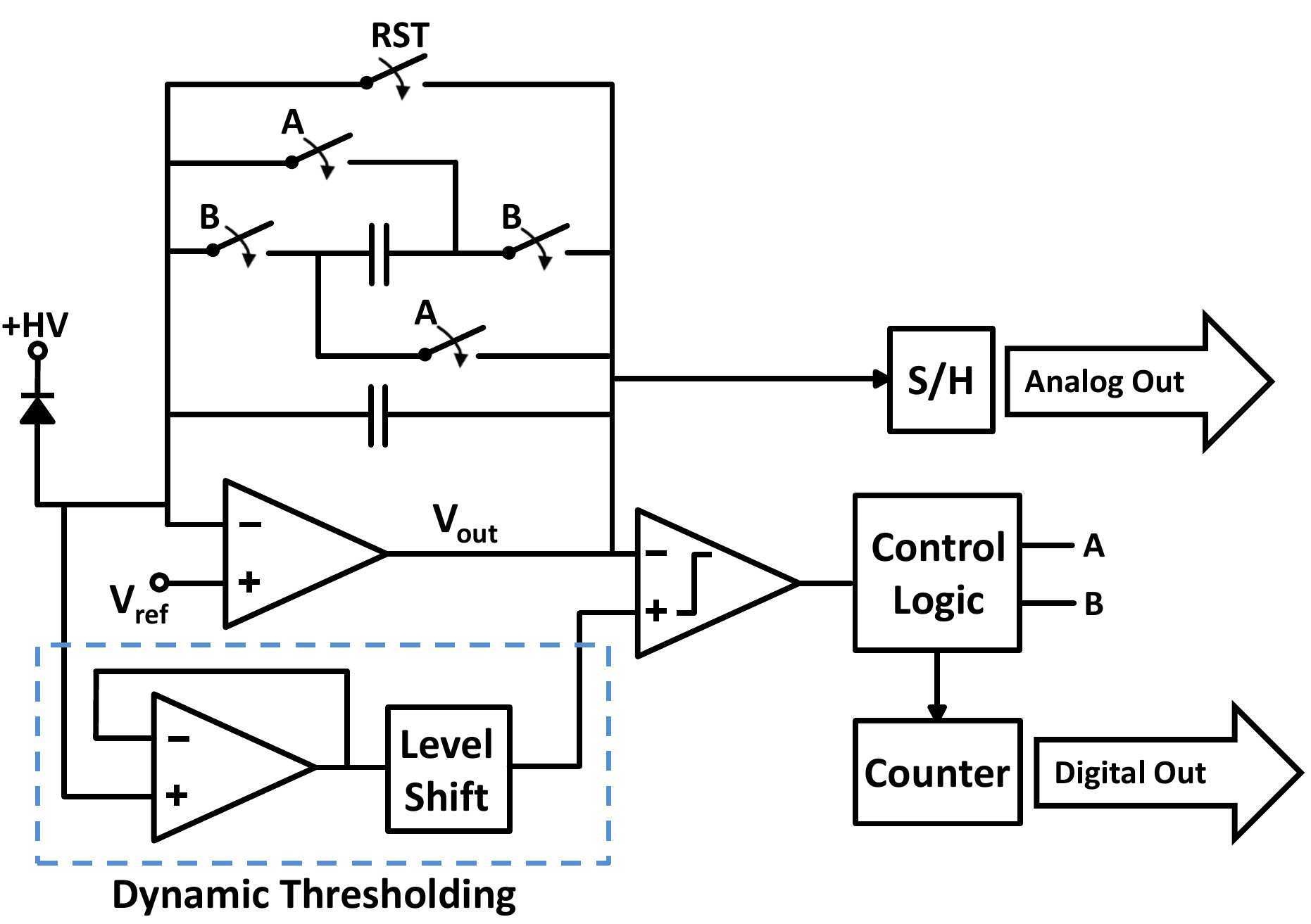}
\caption{Simplified capacitor flipping pixel schematic. Dynamic thresholding circuitry is enclosed in the dotted box. An adaptive gain scheme was implemented in the most recent fabrication.}\label{capflip}
\end{figure}

The third pixel front-end fabricated and tested relies on a charge removal method based on the flipped capacitor filter described in \cite{Manghisoni2015}. The capacitor flipping pixel integrator uses two equally sized integration capacitors connected in parallel, depicted in figure \ref{capflip}. One of the two integration capacitors is connected via a network of CMOS switches so as to allow the capacitor's polarity in the circuit to be reversed. Adaptive gain was not incorporated in the initial fabrication and is not shown in figure \ref{capflip}, but it was implemented in the 16x16 pixel array test ASIC described in section VII.  

Referring to figure \ref{capflip}, the pixel begins integration with the switches labeled “A” closed and the switches labeled “B” open. Charge is integrated onto both equal-sized capacitors in parallel. When the integrator output crosses the comparator threshold, the comparator fires. This activates the control logic which opens switches “A”. After a brief delay to prevent shorting the integration capacitors, switches “B” are closed. This reverses the orientation of half of the integration capacitance. As a result, integrated positive charge on each capacitor neutralizes the negative charge on the other capacitor. Thus, the integrator is effectively reset, and can continue integrating photocurrent. Subsequent flipping occurs as needed. Connections to the flipping capacitor are always broken before new connections are made.

As with each other charge removal circuit, $\Delta Q$ depends on the front-end voltage of the integrator. To reduce the error introduced by changes in the front-end voltage due to large current spikes, a dynamic thresholding circuit was devised. In its present implementation, the pixel comparator can use an external reference voltage to define the threshold at which capacitor flipping is initiated, or it can be set dynamically. In the dynamic case, a level-shifted copy of the front-end voltage is used to set the capacitor flipping threshold voltage. This ensures that there is a specific voltage across the integration capacitance at the time of the comparator firing, precisely the level shift voltage. The dynamic thresholding block level schematic is depicted in figure \ref{capflip}.

\section{Constant Current Integration}
The three pixel front-ends were fabricated with several means of injecting a test charge to emulate an input x-ray signal. A PMOS current source in each pixel provided simple functionality tests. For higher currents and quantitative results, a copy of each pixel with a probe pad attached to its input was included in fabrication. Current was injected into pixels through a tungsten needle with a 10k$\Omega$ resistor between the needle and an external current source. The current was generated and regulated by a Keithley 2400 Sourcemeter. Output signals were buffered off chip to a DPO7254C Tektronix oscilloscope. 

Parasitic capacitance of the needle probe was estimated to be $\sim$10pF. This estimate is based on the change of $V_{out}$ in the MM-PAD 2.0 pixel during a charge removal event. The integrating amplifier output voltage jumps during charge removal to maintain $V_{ref}$ on the front-end. The jump is smaller with the needle contacting the front-end because the needle's parasitic capacitance reduces the charge transfer efficiency of the integrating amplifier, i.e. charge removal pulls some charge from the parasitic capacitance rather than the integration capacitance. This parasitic capacitance is significantly larger than the contribution expected from a bump bonded sensor, and therefore pixel performance under single pulse current injection would not accurately represent a hybridized pixel's performance integrating x-ray pulses. However, the performance of a pixel with a constant current input is still indicative of its ability to integrate high sustained x-ray flux.

\section{Results} 
\subsection{Linearity of Integration}
\begin{figure}[!t]
\centering
\includegraphics[width=3.3in]{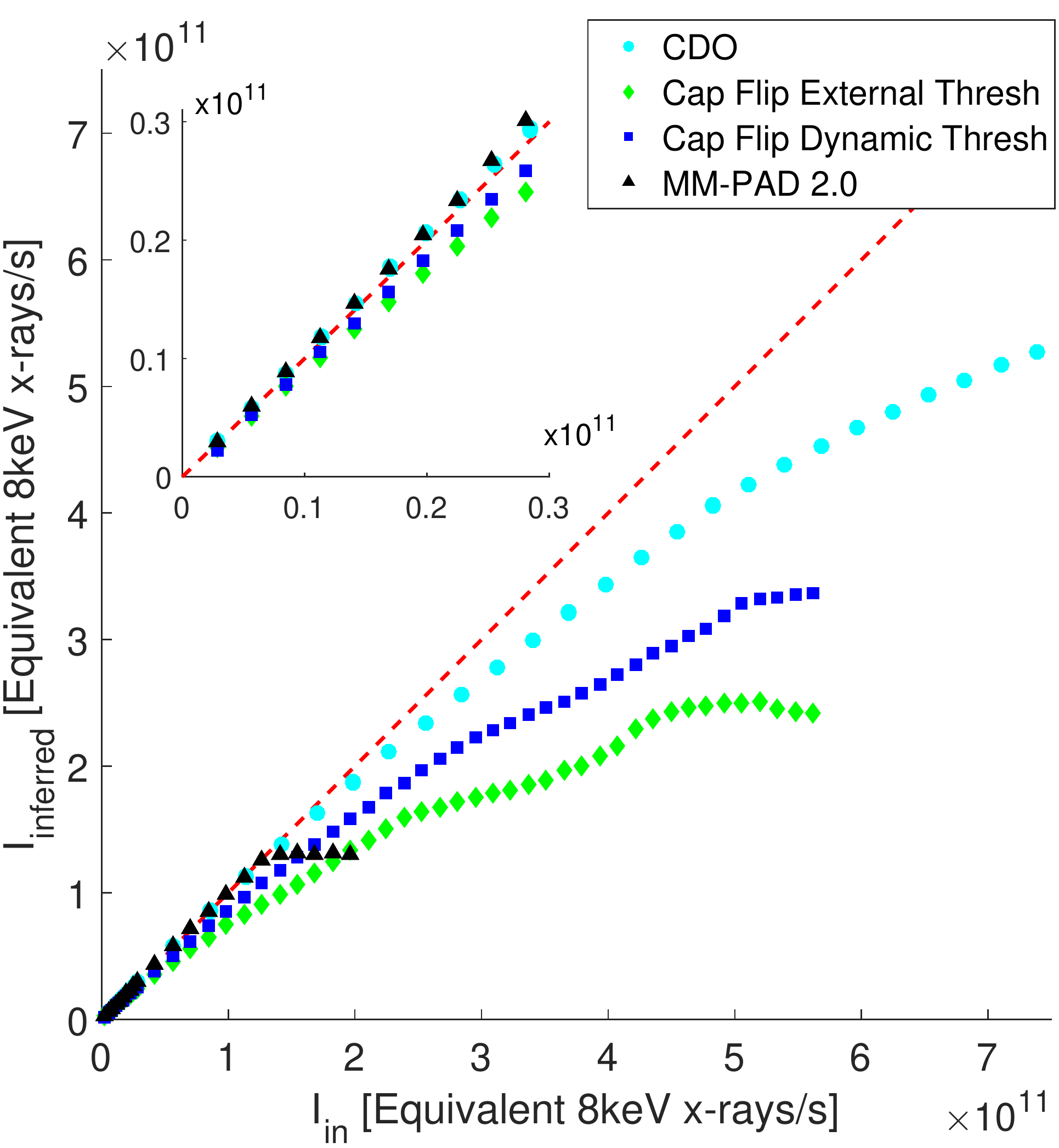}
\caption{Inferred input currents based on pixel outputs versus actual input current. The dotted line represents an ideal response (inferred input equals actual input). The charge dump oscillator is plotted with circles, the MM-PAD 2.0 with triangles, the externally thresholded capacitor flipping pixel with diamonds, and the dynamically thresholded capacitor flipping pixel with squares. Input and inferred current values are converted to the number of 8 keV x-rays absorbed in silicon per second which would produce an equivalent photocurrent. Inset: Magnification of the same data.}\label{iinferred}
\end{figure}

To evaluate the linearity of integration for each pixel architecture, pixel output was monitored with a constant current input. The measured output, charge removal frequency, was  multiplied by nominal $\Delta Q$ values (the quantity of charge removed per removal execution) to calculate an inferred input current. This inferred input current can then be compared to the actual, known input current. These values are plotted against each other in figure \ref{iinferred}. To measure the charge removal frequency, buffered charge removal control signals were measured on an oscilloscope, and edge finding algorithms were used to determine the time of each charge removal cycle. Linear fits to the time of each charge removal versus the number of charge removals preceding it yielded the charge removal period as the slope of the fit. From this, frequency and uncertainty in the determination of the frequency were extracted. Variations in frequency between traces at a given input current were larger than the uncertainty in the determination of the frequency in a single trace, but both measures of uncertainty are smaller than the data points plotted in figure \ref{iinferred}. Voltages used in the calculation of charge removal quantities were taken at their nominal values, e.g., $V_{front-end}$ was taken as $V_{ref}$, which is set externally. Throughout this section, current is specified in units of equivalent 8 keV x-rays/s. This is the flux of 8 keV x-rays that, when absorbed in a reverse biased silicon diode, would produce an equivalent photocurrent.

The MM-PAD 2.0 results are shown as triangles in figure \ref{iinferred}. Good performance is seen with inputs up to $1.3x10^{11}$ 8 keV x-rays/s equivalent. The inferred current measurement eventually plateaus, indicating that the pixel oscillator is operating at its maximum frequency. These values are consistent with simulation. Deviation from linearity is likely a result of process variation in charge removal capacitor size and $V_{front-end}$ not being held precisely at $V_{ref}$. This can be calibrated.

The CDO results are shown as circles in figure \ref{iinferred}. At low input currents, the inferred input is greater than the actual input, which implies that $\Delta Q$ is less than what is expected based on the value of $V_{ref}$. This could be a result of process variation in capacitor size. Alternatively, incomplete charge removal may occur because signals in the ring oscillator propagate quickly compared to time constants associated with the charge dump process. However, if the dump is repeatable it can potentially be calibrated.

As the input current increases, the CDO's inferred input current drops below the actual input current. In this regime, above $~10^{11}$ x-rays/s, the quantity of charge removed per charge removal execution exceeds the expected value. A likely cause of this error is a significant rise in the pixel front-end voltage above $V_{ref}$. This would cause more integrated charge to be removed from the integration node than intended.

Capacitor flipping pixel data was taken with both dynamic thresholding and a fixed, external threshold. These data are plotted in figure \ref{iinferred} as squares and diamonds, respectively. Some error in externally thresholded operation is a result of the integrator front-end voltage drifting upwards with higher input currents. This drift was observed directly in testing. The decrease of inferred current above $5x10^{11}$ 8 keV x-rays/s equivalent input in the externally thresholded case is likely a result of the front-end voltage being pushed outside of the integrating amplifier's range of optimal operating conditions. 

\begin{figure}[!t]
\centering
\includegraphics[width=3.25in]{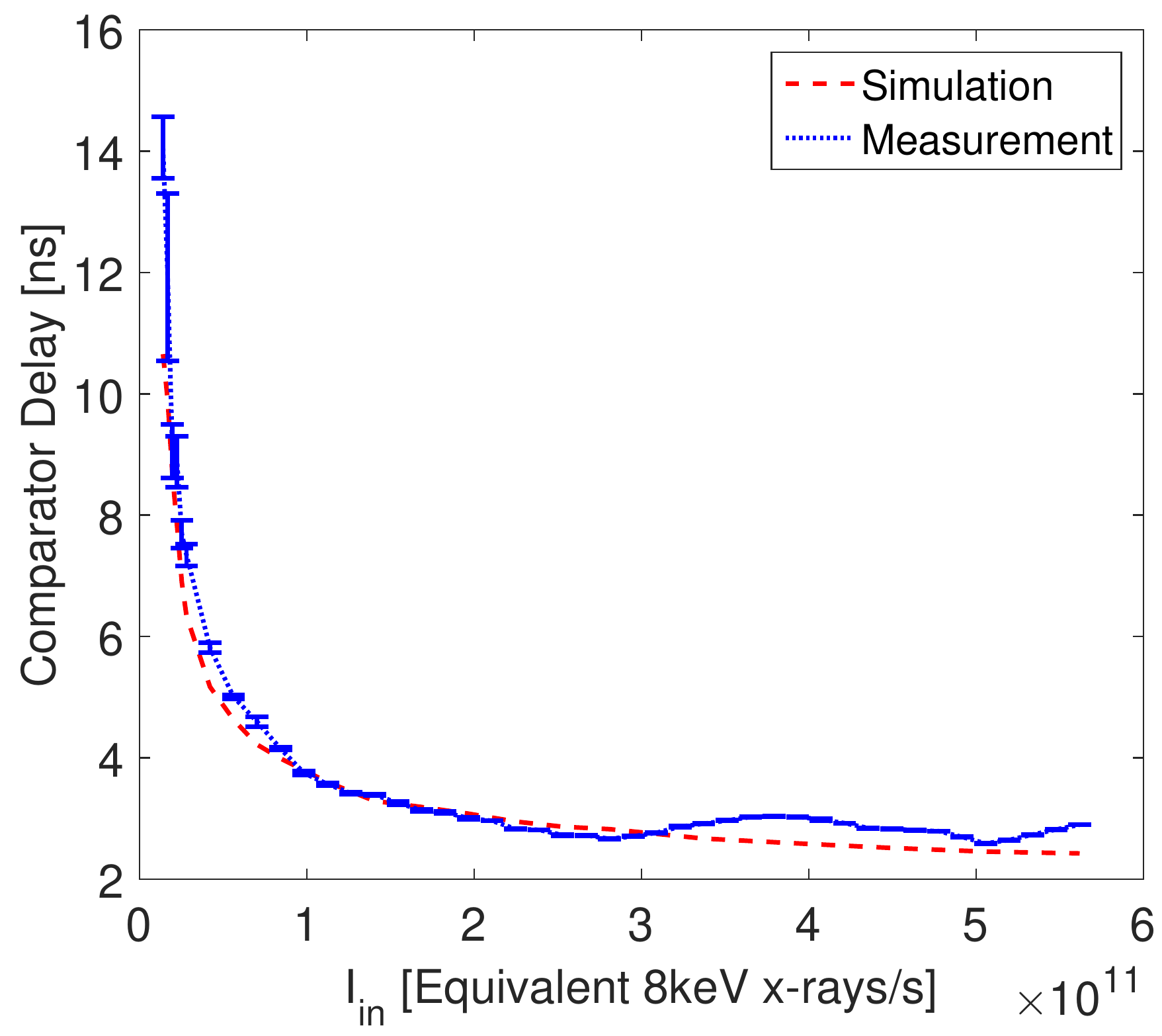}
\caption{Measured comparator delays from the capacitor flipping pixel with dynamic thresholding are plotted. Measured values assume that all deviations from linearity in the capacitor flipping pixel's output are a result of charge integrated during switching delays. Values from simulation are plotted as a dotted line.}\label{compdelay}
\end{figure}

A clear improvement in performance is seen with the dynamic thresholding enabled. However, there is still substantial error in the input reconstruction: less than 70$\%$ of the input is accounted for above $3.5\times{}10^{11}$ x-rays/s equivalent input. This error can be explained by noting that the quantity of charge neutralized per capacitor flip, $\Delta Q$, depends on the voltage across the integrator, not just the front-end voltage as in the other pixel designs. This means that any delay between when the capacitor should be disconnected and when it actually does disconnect can introduce error. Specifically, if photocurrent continues to be integrated during this delay, the output voltage of the integrator will continue to drop and the charge neutralized will be greater than anticipated.

From these data we can extract the error per capacitor flip. This is the difference between actual and inferred input currents divided by the frequency of capacitor flipping. Put another way, this is the charge removed per capacitor flip beyond what is expected based on the value of the level shift. Simulations of the comparator employed in this particular pixel front-end show that its firing delay varies with input falling edge slope, or equivalently in this case, input current. If we assume that all of this deviation from linearity is a result of photocurrent accumulation during the switching delay, dividing the error per capacitor flip by input current yields a measurement of this delay. Figure \ref{compdelay} plots the measured delays (assuming that all error comes from the delay) on top of the switching delays from simulation, both as functions of input current. 

The measurement appears to follow the simulated values. This highlights a problem inherent to the capacitor flipping charge removal design. Any delay between when the capacitor should flip and when it actually does creates a window in which integrated charge will not be accounted for. Some of the calculated error may be due to an offset in the comparator threshold, but this potential for error is ultimately inherent to the design.

\subsection{Power Consumption}

\begin{table}[!t]
\renewcommand{\arraystretch}{1.3}
\caption{Pixel Average Power Consumption From Simulation}
\label{table_summary}
\centering
\begin{tabular}{|c|c|c|c|c|}
\hline
&& MM-PAD 2.0 & CDO & Cap Flip\\
\hline
\multirow{3}{.8in}{\parbox{.8in}{\centering Active Power\newline(Integrating $10^{11}$ 8 keV x-rays/s)}} & Analog & 146\textmu{}W & 52.7\textmu{}W & 158\textmu{}W\\
& Digital & 79.8\textmu{}W & 130\textmu{}W & 27.2\textmu{}W\\
& \bf{Total} & \bf{226\textmu{}W} & \bf{183\textmu{}W} & \bf{185\textmu{}W}\\
\hline
\multirow{3}{.8in}{\parbox{.8in}{\centering Quiescent Power}} & Analog & 102\textmu{}W & 52.7\textmu{}W & 194\textmu{}W\\
& Digital & 3.63nW & 775nW & 0.563nW\\
& \bf{Total} & \bf{102\textmu{}W} & \bf{53.5\textmu{}W} & \bf{194\textmu{}W}\\
\hline
\end{tabular}
\end{table}

While maximizing the input range of pixels, it is essential to keep power consumption manageable. Power consumption was measured in simulation for each pixel substructure and is listed in Table \ref{table_summary}. Performance of the pixels in simulation was commensurate with their measured performance. These figures have been deemed suitable for scaling of pixel designs to full arrays based on the estimated cooling power available for a full array. 

The capacitor flipping pixel exhibits decreased analog power consumption under high loads. This is because the integrating amplifier in this pixel is a class AB amplifier, and after triggering a charge removal event, it is not required to slew back up to achieve its quiescent voltage. Instead, integrated charge is transferred to the integrator output by the capacitor flipping, and voltage is restored with minimal current supplied by the amplifier. This is not the case in the other pixel architectures.

\section{Potential Application to XFELs}
Charge removal circuitry is a valuable tool to extend measurable signal levels when a large, sustained photocurrent is generated in a pixel. XFELs produce exceedingly bright x-ray pulses with durations on the order of femtoseconds. It would seem that charge removal is ill-suited to the problem of integrating large XFEL pulses because no circuitry can respond on femtosecond time scales.

However, the peak photocurrent generated in pixels by XFEL pulses is not quite as dire as femtosecond x-ray pulse durations suggest. While an entire XFEL pulse reaches a detector in the span of femtoseconds, drift, diffusion, and the plasma effect cause the resulting photocurrent to take significantly longer to arrive at pixel integration nodes \cite{Becker2010b}. 

The plasma effect refers to the case when a sufficiently large number of electron-hole pairs are generated in a sufficiently small volume of the photo-sensor so as to behave like a plasma cloud. X-ray pulses of sufficient intensity can lead to the plasma effect in silicon diodes. The electron-hole plasma expels the photodiode electric field which would ordinarily separate charge carriers and bring them to respective sensor terminals. Instead, the surface of the plasma cloud is wicked away by the expelled electric field while the interior of the plasma remains relatively shielded. This slows down the accumulation of photocurrent at the pixel integration nodes. 

\begin{figure}[!t]
\centering
\includegraphics[width=3.3in]{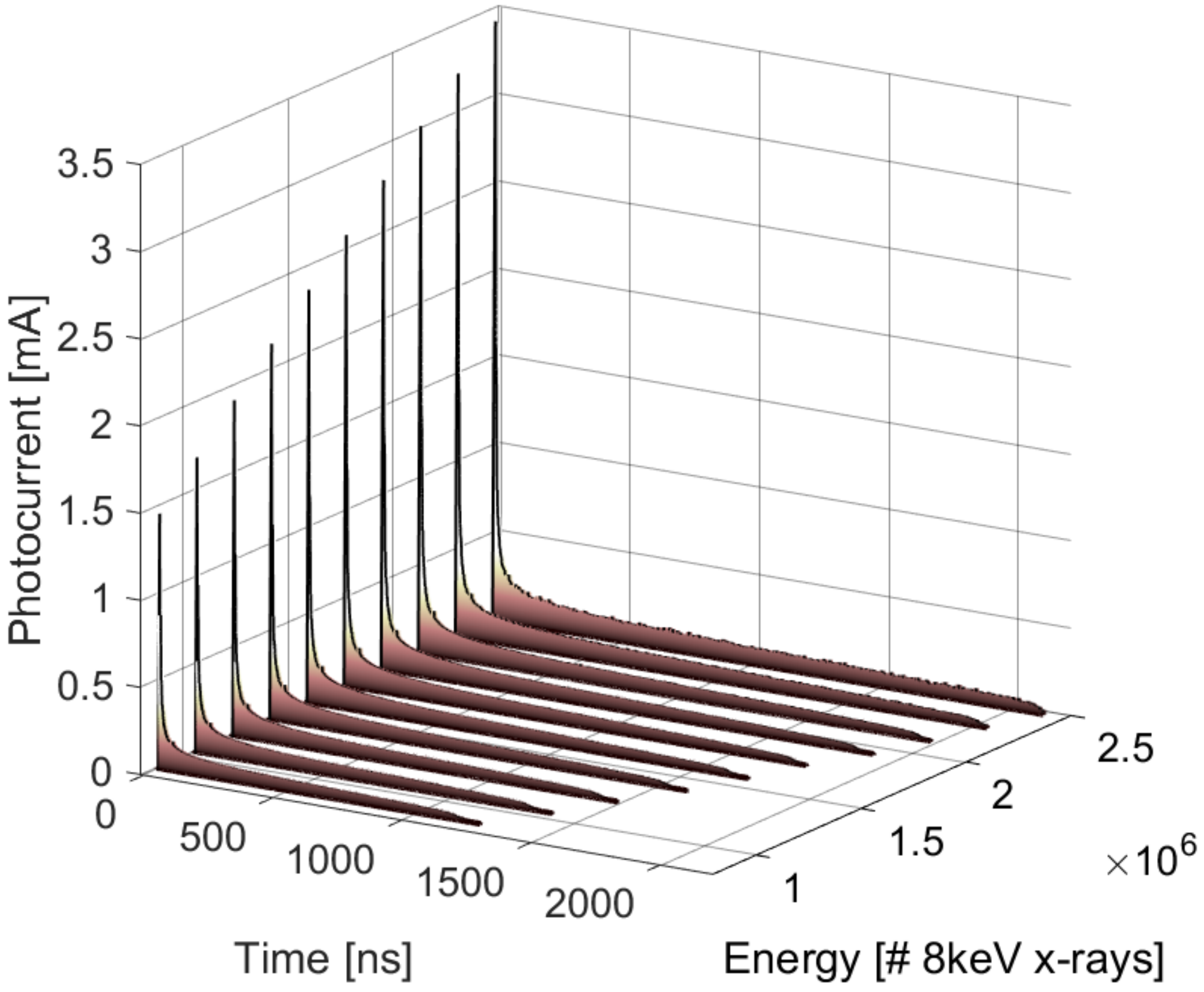}
\caption{Averaged photocurrent traces from 950nm picosecond laser pulses focused to 6\textmu{}m incident on a 500\textmu{}m thick silicon diode biased at 200V. Measurements were taken with the apparatus described in \protect\cite{Weiss}.}\label{traces}
\end{figure}

To better understand this process, and to assess the prospect of operation at XFELs, we have utilized the transient current technique to measure photocurrent transients from a pixelated silicon diode illuminated by a focused infrared laser, as described in \cite{Weiss}. Laser wavelengths were chosen to match the attenuation length of x-ray photons in silicon. For example, 950 nm infrared light has the same attenuation length in silicon at room temperature as 8 keV x-rays. As a result, absorption of an intense 950 nm laser pulse in a silicon diode produces an electron-hole pair distribution in the sensor which is similar to what we might expect from a pulse of 8 keV x-rays.

Figure \ref{traces} is a plot of averaged photocurrent transients collected from a silicon diode similar to the diode used in the MM-PAD. Generally speaking, photocurrent arrives in two phases: an initial spike arising largely from induced current \cite{he2001review} followed by a long tail as charge carriers drift to the sensor terminals. A strategy to integrate pulses of this nature might involve adaptive gain handling the integration of the initial photocurrent spike and charge removal circuitry handling the drawn out tail. A small scale detector prototype, featuring the pixels described above, with a bump bonded silicon photodiode will be used to image these laser pulses and assess the viability of this strategy.

\section{Summary and Future Work}

The performance of the pixel substructures discussed above demonstrates that each is capable of integrating large quantities of photocurrent. The MM-PAD 2.0 exhibits robust performance up to the design goal of $10^{11}$ 8 keV x-rays/pixel/s. However, to handle larger signals a new oscillator would be required. By coupling the oscillator and charge dump circuitry so closely, the CDO demonstrates a viable strategy for extending sustained count rates further than the MM-PAD 2.0, but concerns about its pixel-to-pixel variation require further investigations which were not performed in this study. The capacitor flipping pixel exhibits a systematic deviation from linearity which is ultimately undesirable in an x-ray pixel for scientific work, but the effectiveness of the dynamic thresholding concept is demonstrated.

The pixel front-ends discussed above have been fabricated as a test ASIC. The ASIC has a 16x16 pixel array with four different pixel designs (the MM-PAD 2.0, the CDO, the flipping capacitor, and a further modification of the MM-PAD 2.0), each with an in-pixel counter, adaptive gain, fully functional readout, and addressable, in-pixel input current sources for functional testing. 

The array will be bump bonded to a 16x16 pixel silicon sensor to test pixel performance with the integration of x-rays and infrared laser pulses. An additional row of pixels have been fabricated with probe pad inputs to allow direct current injection testing as described in this paper. The pixel array will be used to test integration performance at both high and low fluence. Whichever pixel is best able to measure the full signal range will be developed further. Testing of this pixel array detector will begin in late 2016.

\section*{Acknowledgments}

This research is supported by the U.S. National Science Foundation and the U.S. National
Institutes of Health/National Institute of General Medical Sciences via NSF award
DMR-1332208, U.S. Department of Energy Award DE-FG02-1 0ER46693 and DE-SC0016035,
and the W. M. Keck Foundation. The MM-PAD concept was developed collaboratively by
our Cornell University detector group and Area Detector Systems Corporation, Poway, CA, USA.

\bibliographystyle{IEEEtran}
\bibliography{Weiss_HDRPAD}

\begin{thebibliography}{10}
\providecommand{\url}[1]{#1}
\csname url@samestyle\endcsname
\providecommand{\newblock}{\relax}
\providecommand{\bibinfo}[2]{#2}
\providecommand{\BIBentrySTDinterwordspacing}{\spaceskip=0pt\relax}
\providecommand{\BIBentryALTinterwordstretchfactor}{4}
\providecommand{\BIBentryALTinterwordspacing}{\spaceskip=\fontdimen2\font plus
\BIBentryALTinterwordstretchfactor\fontdimen3\font minus
  \fontdimen4\font\relax}
\providecommand{\BIBforeignlanguage}[2]{{%
\expandafter\ifx\csname l@#1\endcsname\relax
\typeout{** WARNING: IEEEtran.bst: No hyphenation pattern has been}%
\typeout{** loaded for the language `#1'. Using the pattern for}%
\typeout{** the default language instead.}%
\else
\language=\csname l@#1\endcsname
\fi
#2}}
\providecommand{\BIBdecl}{\relax}
\BIBdecl

\bibitem{Carini2012}
\BIBentryALTinterwordspacing
G.~Carini, P.~Denes, S.~M. Gruner, and E.~Lessner, ``Neutron and x-ray
  detectors,'' USDOE Office of Science (SC)(United States), Tech. Rep., 2012.
  [Online]. Available:
  \url{http://science.energy.gov/~/media/bes/pdf/reports/files/NXD_rpt_print.pdf}
\BIBentrySTDinterwordspacing

\bibitem{Nugent2010}
K.~A. Nugent, ``Coherent methods in the x-ray sciences,'' \emph{Advances in
  Physics}, vol.~59, no.~1, pp. 1--99, 2010.

\bibitem{denes2014pixel}
P.~Denes and B.~Schmitt, ``Pixel detectors for diffraction-limited storage
  rings,'' \emph{Journal of synchrotron radiation}, vol.~21, no.~5, pp.
  1006--1010, 2014.

\bibitem{Giewekemeyer2014}
K.~Giewekemeyer, H.~T. Philipp, R.~N. Wilke, A.~Aquila, M.~Osterhoff, M.~W.
  Tate, K.~S. Shanks, A.~V. Zozulya, T.~Salditt, S.~M. Gruner \emph{et~al.},
  ``High-dynamic-range coherent diffractive imaging: ptychography using the
  mixed-mode pixel array detector,'' \emph{Journal of synchrotron radiation},
  vol.~21, no.~5, pp. 1167--1174, 2014.

\bibitem{Tate2013}
M.~Tate, D.~Chamberlain, K.~Green, H.~Philipp, P.~Purohit, C.~Strohman, and
  S.~Gruner, ``A medium-format, mixed-mode pixel array detector for kilohertz
  x-ray imaging,'' in \emph{Journal of Physics: Conference Series}, vol. 425,
  no.~6.\hskip 1em plus 0.5em minus 0.4em\relax IOP Publishing, 2013, p.
  062004.

\bibitem{Schuette2008}
D.~R. Schuette, ``A mixed analog and digital pixel array detector for
  synchrotron x-ray imaging,'' Ph.D. dissertation, Cornell University, 2008.

\bibitem{Shanks}
K.~S. Shanks, H.~T. Philipp, J.~T. Weiss, J.~Becker, M.~W. Tate, and S.~M.
  Gruner, ``The high dynamic range pixel array detector (hdr-pad): Concept and
  design,'' \emph{AIP Conference Proceedings}, vol. 1741, no.~1, 2016.

\bibitem{Henrich2011}
B.~Henrich, J.~Becker, R.~Dinapoli, P.~Goettlicher, H.~Graafsma, H.~Hirsemann,
  R.~Klanner, H.~Krueger, R.~Mazzocco, A.~Mozzanica \emph{et~al.}, ``The
  adaptive gain integrating pixel detector agipd a detector for the european
  xfel,'' \emph{Nuclear Instruments and Methods in Physics Research Section A:
  Accelerators, Spectrometers, Detectors and Associated Equipment}, vol. 633,
  pp. S11--S14, 2011.

\bibitem{Manghisoni2015}
M.~Manghisoni, D.~Comotti, L.~Gaioni, L.~Lodola, L.~Ratti, V.~Re, G.~Traversi,
  and C.~Vacchi, ``Novel active signal compression in low-noise analog readout
  at future x-ray fel facilities,'' \emph{Journal of Instrumentation}, vol.~10,
  no.~04, p. C04003, 2015.

\bibitem{Becker2010b}
J.~Becker, D.~Eckstein, R.~Klanner, and G.~Steinbr{\"u}ck, ``Impact of plasma
  effects on the performance of silicon sensors at an x-ray fel,''
  \emph{Nuclear Instruments and Methods in Physics Research Section A:
  Accelerators, Spectrometers, Detectors and Associated Equipment}, vol. 615,
  no.~2, pp. 230--236, 2010.

\bibitem{Weiss}
J.~T. Weiss, J.~Becker, K.~S. Shanks, H.~T. Philipp, M.~W. Tate, and S.~M.
  Gruner, ``Potential beneficial effects of electron-hole plasmas created in
  silicon sensors by xfel-like high intensity pulses for detector
  development,'' \emph{AIP Conference Proceedings}, vol. 1741, no.~1, 2016.

\bibitem{he2001review}
Z.~He, ``Review of the shockley--ramo theorem and its application in
  semiconductor gamma-ray detectors,'' \emph{Nuclear Instruments and Methods in
  Physics Research Section A: Accelerators, Spectrometers, Detectors and
  Associated Equipment}, vol. 463, no.~1, pp. 250--267, 2001.

\end{thebibliography}

\end{document}